\documentclass[]{interact}

\usepackage{epstopdf}
\usepackage[caption=false]{subfig}

\usepackage[numbers,sort&compress]{natbib}

\usepackage{algorithm}
\usepackage{algpseudocode}
\usepackage{graphicx}%
\usepackage{multirow}
\usepackage[font={footnotesize}]{caption}

\bibpunct[, ]{[}{]}{,}{n}{,}{,}
\renewcommand\bibfont{\fontsize{10}{12}\selectfont}

\theoremstyle{plain}

\theoremstyle{definition}

\theoremstyle{remark}

\graphicspath{{./figures/4options}{./figures/16options}}

\begin{document}


\title{Neural Networks for Portfolio-Level Risk Management: Portfolio Compression, Static Hedging, Counterparty Credit Risk Exposures and Impact on Capital Requirement}

\author{
\name{Vikranth Lokeshwar Dhandapani\textsuperscript{a}\thanks{CONTACT Vikranth Lokeshwar Dhandapani. Email: vikranthl@iisc.ac.in} and Shashi Jain\textsuperscript{a}\thanks{CONTACT Shashi Jain. Email: shashijain@iisc.ac.in} }
\affil{\textsuperscript{a} Department of Management Studies, Indian Institute of Science, Bangalore - 560012, Karnataka, India; 
}}

\maketitle

\begin{abstract}
In this paper, we present an artificial neural network framework for portfolio compression of a large portfolio of European options with varying maturities (target portfolio) by a significantly smaller portfolio of European options with shorter or same maturity (compressed portfolio), which also represents a self-replicating static hedge portfolio of the target portfolio. For the proposed machine learning architecture, which is consummately interpretable by choice of design, we also define the algorithm to learn model parameters by providing a parameter initialisation technique and leveraging the optimisation methodology proposed in \cite{dhandapani2024bermudan}, which was initially introduced to price Bermudan options. We demonstrate the convergence of errors and the iterative evolution of neural network parameters over the course of optimization process, using selected target portfolio samples for illustration. We demonstrate through numerical examples that the Exposure distributions and Exposure profiles (Expected Exposure and Potential Future Exposure) of the target portfolio and compressed portfolio align closely across future risk horizons under risk-neutral and real-world scenarios. Additionally, we benchmark the target portfolio's Financial Greeks (Delta, Gamma, and Vega) against the compressed portfolio at future time horizons across different market scenarios generated by Monte-Carlo simulations. Finally, we compare the regulatory capital requirement under the standardised approach for counterparty credit risk of the target portfolio against the compressed portfolio and highlight that the capital requirement for the compact portfolio substantially reduces. 
\end{abstract}



\section{Introduction}

Portfolio compression is a post-trade netting and risk mitigation mechanism. It aims to reduce a substantially large portfolio of multiple offsetting derivative contracts to a portfolio of fewer deals and gross notional values. This process proportionately abates counterparty credit risk, bookkeeping tasks, and transaction costs. Importantly, it achieves this without changing the market risk exposure or with the exposure change within an acceptable tolerance limit (refer the European Securities and Markets Authority (ESMA) report on post trade risk reduction services \cite{esma} for detailed definitions of portfolio compression). Risk management has been continuously evolving, as discussed in \cite{dionne2013risk}, and portfolio compression as a risk mitigation technique was widely acknowledged, particularly after it became a part of various financial legislations in the US and European markets. \cite{o2017optimising} numerically analysed a set of multilateral netting algorithms based on exposure minimisation and showed that optimal multilateral netting is an effective counterparty credit risk mitigation technique. \cite{aldasoro2018credit} stated that portfolio compression of the bilateral and multilateral portfolios was a key driving factor for notional amounts of credit default swap (CDS) contracts to fall markedly from USD 61.2 trillion at the end of 2007 to 9.4 trillion in 10 years. \cite{ehlers2019evolution} has highlighted that the portfolio compressed trades steadily increased from 2016 to 2019, which was the leading cause for the average daily turnover of OTC interest rate derivatives to grow more than twofold in the three years. The BIS report by \cite{schrimpf2015outstanding} shows evidence for the decline (prominently in the interest rate and CDS market) in the gross notional of credit, interest rate, commodity, fx, and equity asset class instruments in the OTC derivatives market. The novation of the central counterparty should inflate the trades (and therefore, the notional) as a single transaction between two dealers becomes two transactions with a central counterparty; however, the portfolio compression has reversed this trend and sharply undercut the value of outstanding notional. \\

\noindent Extensive research has focussed on the impact of post-trade risk mitigation techniques on systemic risk, compression of financial networks, and efficient portfolio compression. \cite{veraart2020distress} developed a new model for solvency contagion used to quantify systemic risk in stress tests of financial networks. \cite{veraart2021macroprudential} analysed the consequences of post-trade risk reduction services (portfolio rebalancing and compression) for systemic risk in derivatives markets and provided sufficient conditions for portfolio rebalancing to reduce systemic risk. \cite{veraart2022does} discussed the conditions where portfolio compression could increase and decrease systemic risk. The author has shown that, in most cases, portfolio compression reduces systemic risk. Still, in some cases, firms engaged in compression increased systemic risks and are at risk of default because it might pose large losses to counterparties and lead to contagious default. \cite{amini2022optimal} studied the optimal network compression problem and focussed on objective functions generated by systemic risk measures under shocks to the financial network. \cite{dokuchaev2001optimal} focussed on optimal portfolio selection and compression in an incomplete market. \cite{marquart2014note} detailed the concept of multilateral portfolio compression and proposed steps for the compression of vanilla interest rate swaps. \cite{d2021compressing} have proposed different portfolio compression frameworks to reduce the over-the-counter market size carrying the same net exposure without considering the impact of compression on systemic risk. \\

\noindent Drawing from research outcomes on portfolio replication such as \cite{dembo1999practice}, scenario-based approach (for non-linear instruments like options) is very effective in portfolio replication. Few applications of portfolio replication, as mentioned in \cite{dembo1999practice}, include portfolio compression, portfolio tracking, and static hedging. This perspective views scenarios based portfolio replication as an effective tool for portfolio compression. Few notable researches on portfolio replication include \cite{derman1995static}, where the construction of replicating portfolio of standard options with varying strikes and maturity for hedging target stock options is discussed. \cite{dempster2002dynamic} considered dynamic portfolio replication and portfolio compression using stochastic programming. \cite{carr2014static} proposed a static spanning relation between target options and a continuum of shorter-term options written on the same asset under a single-factor Markovian setting. \cite{bossu2021functional} used the spectral decomposition technique to replicate payoff by a discrete portfolio of special options, which also can be a powerful method for fast pricing large vanilla option portfolios. \\

\noindent There have been concerted efforts by academicians and practitioners to apply artificial intelligence in risk management of financial derivatives in the last decade (\cite{leo2019machine}); in particular, explainable algorithms have gained major attraction to accommodate regulatory model frameworks. \cite{hoencamp2022semi} approached a neural network based semi-static replication approach to efficient hedging and pricing of callable IR derivatives. \cite{lokeshwar2022explainable} proposed an explainable neural network (called as RLNN) to price and statically hedge a single option (inclusive of exotics) by a portfolio of short-term vanilla options with all neural network parameters learned by the back-propagation algorithm. \cite{dhandapani2023data} empirically studied the RLNN algorithm's effectiveness in statically hedging monthly index options using weekly options on the National Stock Exchange, a prominent exchange in India. The analysis considered real-time constraints such as transaction costs, liquidity, and option availability. In \cite{dhandapani2024bermudan}, an enhanced neural network framework (RLNN-OPT) with faster convergence tailored for pricing and static hedging of Bermudan options was presented. The primary distinction between RLNN and RLNN-OPT lies in how the strikes and weights of the constituent options within the static hedge portfolio are learned. In the back-propagation algorithm, RLNN utilizes the Adam optimizer to learn the necessary strikes and weights, while RLNN-OPT employs the Adam optimizer for strikes and linear regression for weights iteratively. However, both these algorithms have been studied in the context of pricing and hedging of individual options.  \\ 

\noindent In reality, risk management at portfolio level is a necessity rather than handling risks at individual option level. This paper can be viewed as an extension of the work by \cite{lokeshwar2022explainable} and \cite{dhandapani2024bermudan} from the realm of individual options to the portfolio level, particularly in the context of risk management. Hence, we capitalize on the optimization techniques employed in the RLNN-OPT neural network architecture and extend its implementation by devising a neural network framework to achieve portfolio compression and hedging. In this paper, the proposed neural network output can be interpreted as the payoff (at expiry) of the smaller portfolio of vanilla options to statically hedge a significantly large portfolio of target vanilla options throughout the span of smaller portfolio. The smaller portfolio of vanilla options will act as the compressed portfolio for the target portfolio. Additionally, each hidden node of the neural network corresponds to a constituent option of the compressed portfolio, while the weights between the hidden layer and output layer correspond to the portfolio weights. The algorithm can perform auto-netting and static hedging with reduced notional if the target portfolio has long and short positions. Nevertheless, the real motivation of portfolio compression in this paper is not towards automating the netting of offsetting positions (which the algorithm manages by design) but to generate a reduced portfolio for the target portfolio of only long positions, which has profound benefits discussed next. \\

\noindent The compressed portfolio can statically hedge the large target portfolio without rebalancing for a considerable time. The algorithm can boost faster pricing of vanilla options (both at time-zero and future simulated time points), which would be helpful in large option books and high-frequency trading. In \cite{lokeshwar2022explainable}, the neural network generates a static hedge portfolio for each exotic option. This hedge portfolio is extensive, equivalent to the number of internal nodes in the neural network. Consequently, when combined at the portfolio level, it can result in a substantially large portfolio of vanilla options used for hedging the target option. The algorithm proposed in this research work addresses this issue by compressing such a large static hedge portfolio into a smaller one. Notably, the algorithm is also a proficient technique to break the longer-term target portfolio into an equivalent shorter-term portfolio, which can be re-invested at the end of every short term to realise the longer-term portfolio payoff and characteristics. Breaking the target portfolio into a shorter-term portfolio can significantly reduce standardised capital for counterparty credit risk under BASEL regulation \cite{biscre52}. The incremental advantage of the proposed methodology is faster evaluation of exposure profiles at different risk horizons. Further, the compressed portfolio can also act as a tracking portfolio and help in different market risk and risk-governance requirements, for example, ongoing monitoring of different market risk characteristics/attributes with reduced effort.

\section{Portfolio Compression Framework}
\label{Portfolio Compression Framework}

\noindent This section illustrates the portfolio compression framework and introduces the notations\footnote{It is to be noted that $\cdot$ is generally used for multiplication, which includes matrix multiplication, and $\odot$ is used for element-wise multiplication between matrices or vectors of the same dimensions. In this paper, the function applied on each element of the vector (or matrix) is notationally mentioned as the function on the vector (or matrix), i.e., for any arbitrary vector $Q=(q_1, q_2, ....q_d)^{\intercal} \in \mathbb{R}^{d}$, $(f(q_1), f(q_2), ..., f(q_d))^{\intercal}$ is recorded as $f(Q)$.} used in this paper. \\

\noindent We assume a complete probability space $(\Omega, \mathcal{F}, \mathbb{P})$,  filtration $\mathcal{F}_t$ for an arbitrary time  $t \in [0,T]$ and an adapted underlying asset process $S_t$, $ \forall t$. The stochastic dynamics of the underlying asset are assumed to follow Geometric Brownian Motion (GBM), and therefore, 
 
\begin{equation}\label{GBMprocessPflCompression}
S_{t} = S_{0} \cdot  exp \left(  \Big( \ r - \frac{\sigma^2}{2} \ \Big) \ t + \sigma \ Z_{t} \right) , 
\end{equation} 

\noindent where, $S_{0}$ is the initial value of the underlying at time $0$, $r$ is the risk-free interest rate, $\sigma$ is the constant volatility and $Z_{t}$ is Brownian Motion. \\

\noindent We aim to compress the target portfolio of $M$ European options (inclusive of call and put options) with constituent option strikes {\Large $\kappa$} $\in \mathbb{R}^{M}$ such that each option starts at time $t_0=0$ and expires at any $t_{i}$, where $i \in \{1, 2, 3, ... \tau \}$, $t_{\tau} = T$ and $t_{i} - t_{i-1} = $ $\Delta t$  $\forall i$;\footnote{In this paper, $\Delta t$ is considered constant for all $i$ for simplicity, but the algorithm is not limited to this condition} The time points $\{t_1, t_2, ... t_{\tau}\}$ are future time points of interest, also called time horizons or risk horizons, in this paper. Therefore, the maturity of each constituent target option could vary from $\Delta t$ to $(\tau \cdot \Delta t)$. The target option portfolio is compressed into a small portfolio of $m$ ($m << M$) European call and put options of maturity $\Delta t$ with strikes $K \in \mathbb{R}^{m}$ and portfolio weights $W \in \mathbb{R}^{m}$ constructed at the beginning of every discrete interval $[t_{i}, t_{i+1}]$, for $i=0, 1, ..., \tau - 1$. Equivalently, for each interval $[t_{i}, t_{i+1}] \ \forall i$, we are interested in hedging the target portfolio by constructing a self-replicating static hedge portfolio, which, if it's also the compressed portfolio, will have both the desirable properties of being a smaller portfolio and also replicating the target portfolio at all $t \in [t_{i}, t_{i+1}]$. In other words,  for a fixed model architecture governed by the neural network hyper-parameters, the compressed portfolio of $m$ options with strike $K$ and portfolio weights $W$  set up at each time $t_i$ by the design of loss function (introduced in Section \ref{Data and Methodology - Chp 5}) will be the optimal static hedge portfolio of the target portfolio till $t_{i+1}$, $\forall \ i $. Therefore, the neural network portfolio acts as both the compressed portfolio and static hedge portfolio for the target portfolio. \\

\noindent The neural network architecture, design of training data and the optimisation methodology are summarised in the following sections.


\subsection{Neural Network Architecture}

\noindent The architecture of the neural network, activation functions, and constraints in this paper bear resemblance to that of \cite{dhandapani2024bermudan}. This similarity arises from the fact that, in \cite{dhandapani2024bermudan}, we construct the static hedge portfolio between consecutive exercise dates for Bermudan options,  while in this paper, we build a static hedge portfolio for a target portfolio of vanilla options. In both scenarios, the instrument type of static hedge portfolio constituents remains the same, resulting in a similar neural network structure to maintain the intended interpretation. \\

\noindent Let us consider the value of the target portfolio as $V(t, S_t)$ at any arbitrary time $t$, where each constituent option is priced using Black-Scholes pricing model introduced by \cite{black1973pricing}. For each $t$ $ \in \{t_1,t_2, ...., t_{\tau} \}$, we are interested in building a compressed portfolio starting at time $t - \Delta t$ and expiring at $t$ with the compressed portfolio's payoff $P(t, S_t)$ at maturity, such that, $||V(t, S_t) - P(t, S_t)||^{2}$ is minimised, where $||.||$ is the L2-norm. This is accomplished by building a neural network model (at each time $t$ with respect to the filtration $\mathcal{F}_{t - \Delta t}$) which inputs underlying asset price and predicts target option portfolio value denoted by $NN(t): S_t \rightarrow P(t, S_t)$, where $P(t, S_t)$ is the predicted value of the target portfolio value $V(t, S_t)$. The output of the proposed neural network $NN(t)$ can be interpreted as the value of the compressed portfolio at time $t$; in other words, the weighted sum of the payoff of $m$ constituent vanilla call and put options expiring at $t$.   \\

\noindent The neural network architecture includes three layers. The first layer has a single node that inputs and outputs the underlying asset price to the hidden layer. The hidden layer has $m$ nodes (corresponding to the $m$ constituent call and put options of the compressed portfolio) with bias term and Rectified Linear Unit (ReLU) activation function in each node. The neural network weights between the input and hidden layers are kept constant (with values $+1$ and $-1$ for the call and put option nodes in the hidden layer). The bias term of each hidden node corresponds to the strike $k_i \ge 0$ (where,  $i = 1, 2, ..., m$) of each constituent option in the compressed portfolio and $K = (k_1, k_2, ..., k_m)^{\intercal}$ $\in$ $\mathbb{R}^{m}$. Further, the activation function of the call option and put option nodes correspond to their payoffs, preserving the ReLU functional form. Consequently, the output of each node is consistent with the payoff of a call or put option $\phi_{i_{cp}}(S_{t}, k_i) = max\big( i_{cp} \cdot (S_{t} - k_i), 0 \big)$, where, $i_{cp} = +1$ for call options and $-1$ for put options. Let $I_{cp} = (i_{cp})^{\intercal}_{i=1,2,..m} \in \mathbb{R}^{m}$ be call/put indicator vector or equivalently the constant neural network weights between the first layer and hidden layer. Therefore, the output payoff vector of the hidden layer can be represented as $\phi(S_{t}, K) = \big(\phi_{i_{cp}}(S_{t}, k_i) \big)_{i=1, .., m}^{\intercal} \in \mathbb{R}^{m}$. The third layer (output layer) has one node which inputs $W^{\intercal} \phi(S_{t}, K)$, where, $W = (w_1, w_2, ..., w_m)^{\intercal} \in \mathbb{R}^{m}$ are the neural network weights (between the hidden layer and output layer) and outputs $P(t, S_t)$. The neural network weights $W$ can be interpreted as portfolio weights corresponding to each constituent option of the compressed portfolio. 

\subsection{Data and Methodology}
\label{Data and Methodology - Chp 5}

There are four topics briefly discussed in this section: training data, loss function, parameter initialisation and optimisation methodology. The optimization methodology presented in \cite{dhandapani2024bermudan} is leveraged and harnessed for portfolio-level data, resulting in effective portfolio compression. This section also presents different steps in the portfolio compression algorithm. 

\begin{itemize}
    \item {\bf Training data \bf}: \\
    The data for training the neural network is obtained by simulating $N$ paths of the underlying asset using the GBM process as defined in Equation \ref{GBMprocessPflCompression} at all time points of interest $t = t_1, t_2, ...., t_{\tau}$. At each simulated path $\omega_j \in \Omega$ (where, $j = 1, 2, ....., N$) of the underlying $S_{t}$, the target portfolio is valued across all time horizons, and the neural network $NN(t)$ is built as a function between underlying price as the independent variable and target portfolio value as the response variable. In other words, $\Big(S_t(\omega_j), V \big(t; S_{t}(\omega_j) \big) \Big)_{j=1}^{N}$ is the feature and response variable data pair for training the neural network $NN(t)$. \\
    
    \item {\bf Loss function \bf}: \\
    The loss function $L(t; W, K)$ is given as: \\
    
    $L(t; W, K)$ $=$ $\frac{1}{2} \cdot ||Y(t) - X(t;K) W||^{2}$,\\
    
    \noindent where, \\
    $Y(t) = \Big( V \big(t; S_{t}(\omega_j) \big) \Big)^{\intercal}_{j=1, 2, ..., N} \in \mathbb{R}^{N}$ is the target the portfolio value, \\
    $X(t;K) = \Big( \phi \big(S_{t}(\omega_j), K \big) \Big)^{\intercal}_{j=1, .., N} \in \mathbb{R}^{N \times m}$ and \\
    $X(t;K)W \in \mathbb{R}^{N}$ is the neural network output or the compressed portfolio value. \\

    \noindent Therefore, by training each neural network $NN(t)$ to minimise $||Y(t) - X(t; K) W||^{2}$, we can determine the constituent $m$ option strikes and their portfolio weights of the optimal compressed portfolio which has to be set up at time $t - \Delta t$ and will also act as the compressed static hedge portfolio for the target portfolio from time $t - \Delta t$ to $t$. The compressed portfolio can be valued at any time $t$ as a weighted sum of each constituent option value, priced by the Black-Scholes pricing model. \\
        
    \item {\bf Parameter Initialisation \bf}: \\
    The parameter vector $K$ is first chosen as equidistant $m$ strikes between deep out-of-the-money (OTM) strike ($50\%$ moneyness) and deep in-the-money (ITM) strike ($150\%$ moneyness), where moneyness is defined as the ratio of strike and spot (Spot/Strike for call and Strike/Spot for put) at time $t_0$. It is to be noted that if $m_c$ and $m_p$ (such that $m = m_c + m_p$) are the number of call and put option nodes of the hidden layer, respectively, $m_c$ and $m_p$ equidistant strikes are chosen separately between deep OTM and ITM strikes and together considered as initial strike vector $K(0)$ to represent corresponding biases of call and put option hidden nodes. At time $t$, by fixing the $m$ initialised strikes, $W$ which minimises $||Y(t) - X(t;K) W||^{2}_{K=K(0)}$ is used as the initial weight vector $W(0)$ of the neural network $NN(t)$. This optimal initialisation is achieved by linear regression (least-squares method without a constant term) between $\phi(S_{t}, K)$ as regressor variables and $V(t, S_t)$ as the response variable and the regression coefficients are considered as optimal initialisation for $W$. \\

    \item{\bf Optimisation \bf}: \\
    The neural network parameters are learned by the back-propagation algorithm (as explained in \cite{bishop1995neural}) with the  optimisation methodology initially introduced in \cite{dhandapani2024bermudan} to iteratively reach the local/global minima.  At each iteration, with a randomly selected fixed batch size of training data (with replacement), initially, the loss function $L(t; W, K)$ is minimised by optimising $K$ for a fixed $W(l)$ using Adaptive Moment Estimation (Adam) optimisation (introduced by \cite{kingma2014adam}) to obtain the next optimal strike vector $K (l+1)$. Finally, by fixing $K(l+1)$, the loss function $L(t; W, K(l+1))$ is minimised by determining optimal $W(l+1)$ vector using linear regression (least-squares method) between payoff vector $\phi(S_{t}, K(l+1))$ and target portfolio value $V(t, S_t)$. The iterations continue till all epoch runs are complete or if the stopping criteria\footnote{If the absolute mean error difference (across all simulated paths at each time horizon) between two consecutive steps, i.e., absolute mean of $\big[L\big(t; W(l+1), K(l+1)\big) - L\big(t; W(l), K(l)\big)\big]$ is less than $10^{-8}$ for ten times in a row.} is reached. \\
    
    \noindent In contrast to the approach taken in \cite{dhandapani2024bermudan}, which primarily centered on developing optimization methodologies for achieving faster convergence, our main emphasis here lies in analyzing the potential of the proposed neural network setup to enhance implementation from individual options to portfolio-level risk management and assessing their performance.\\

\end{itemize}

\begin{algorithm}
\caption{Portfolio Compression Algorithm} \label{algo_portfolio_compression}
\begin{algorithmic}[1]

\State Setup the target portfolio information ($M$, {\Large $\kappa$}), market data ($S_0$, $r$, $\sigma$), the neural network model hyper-parameters ($I_{cp}$, $m$) and optimisation constant parameters relevant to Adam optimiser such as learning rate. 
\State Generate $S_{t} (\omega_j)$ for $j=1, ..., 5000$ using Geometric Brownian Motion defined in Equation \ref{GBMprocessPflCompression} at each time $t \in \{t_1, t_2,..., t_{\tau}\}$.
\State Value the target portfolio ($V(t, S_t(\omega_j) \ \forall j$) at all time points of interest $t \in \{t_1, t_2,...., t_{\tau} \}$.
\State Initialise the parameters $K(0)$ and $W(0)$ of the portfolio compression neural network $NN(t)$ for each time $t$. 
\State Fit the neural network $NN(t)$ at each $t$ as a model between $S_t$ and $V(t, S_t)$ and learn the portfolio weights and constituent option strikes of the compressed portfolio for each time interval $(t-\Delta t,t)$ $\forall t \in \{t_1, t_2,....t_{\tau}\}$.

\end{algorithmic}
\end{algorithm}

\section{Analysis and Results}
\label{Analysis and Results}

In this section, we conduct numerical experiments to analyse the performance of the proposed portfolio compression approach and present the observations. Before delving into a detailed analysis, we introduce the sample test portfolio and market data, and provide a summary of the performed analysis. 

\subsection{Test Portfolio and Market Data}
\noindent We assume a long target portfolio of 10,000  options with 5,000 call and 5000 put options. The strikes of call and put options are selected equidistantly between $80\%$ and $120\%$ moneyness. The maturity of each target option is randomly selected from the discrete Uniform probability mass function $P_{unif}(U): U \rightarrow 0.25$, for each $U \in \{0.25, 0.5, 0.75, 1\}$. We initialise the compressed portfolio strikes and weights as discussed in Section \ref{Data and Methodology - Chp 5} with $m=16$, i.e., a compressed portfolio of $16$ options ($8$ calls and $8$ puts). We also consider a case of compressed portfolio with four constituent options, i.e. $m=4$. \\

\noindent The following market data is considered for the analysis: $S_0 =1$ without loss of generality, $r=0.05$ and $\sigma=0.3$. We assume $5000$ paths for the simulation of the underlying from $t_0 = 0$ to the time points of interest $\{t_1=0.25, t_2=0.5, t_3=0.75, t_4=T=1\}$. The neural network model $NN(t)$ $\forall t \in \{t_1, t_2, t_3, T\}$ is developed using the simulated training data and the model performance is evaluated with the validation data generated by independent simulations with different initial seed as regards training data. \\

\noindent At each time $t \in \{0.25, 0.5, 0.75, 1\}$, we train the neural network $NN(t)$ with 100 epochs of the iterative optimisation algorithm defined to learn the portfolio weights and strikes of the compressed portfolio for each interval $[t-0.25, t]$.

\subsection{Summary of Analysis}
The analysis performed in this paper are summarised below:

\begin{enumerate}
    \item {\bf Model Convergence \bf}: The model error convergence and evolution of neural network parameters ($K$ and $W$) across the epochs at each time horizon are analysed. The model error at a specific time horizon is the absolute mean difference across the simulated paths between the target and compressed portfolio value.
    \item {\bf Benchmarking \bf}: The distribution of PV and Greeks (delta, gamma, and vega) across all simulated paths at each time horizon, along with exposure profiles generated by risk-neutral and real-world measures, is benchmarked for the compressed and target portfolios. The following are the selected attributes to benchmark:
        \begin{itemize}
            \item PV Distribution 
            \item Exposure Benchmarking (risk-neutral and real-world measure)
            \item Greeks Benchmarking
        \end{itemize}
    \item {\bf Standardised Capital \bf}: The standardised counterparty credit risk capital (under BASEL norms) for sample target portfolios and compressed portfolios are evaluated and evidence of reduced capital requirement is highlighted.
\end{enumerate}

\subsection{Model Convergence}
\label{Model Convergence}

\begin{figure}[!htb]
\begin{center}
\includegraphics[width=0.7\textwidth, height=3in]{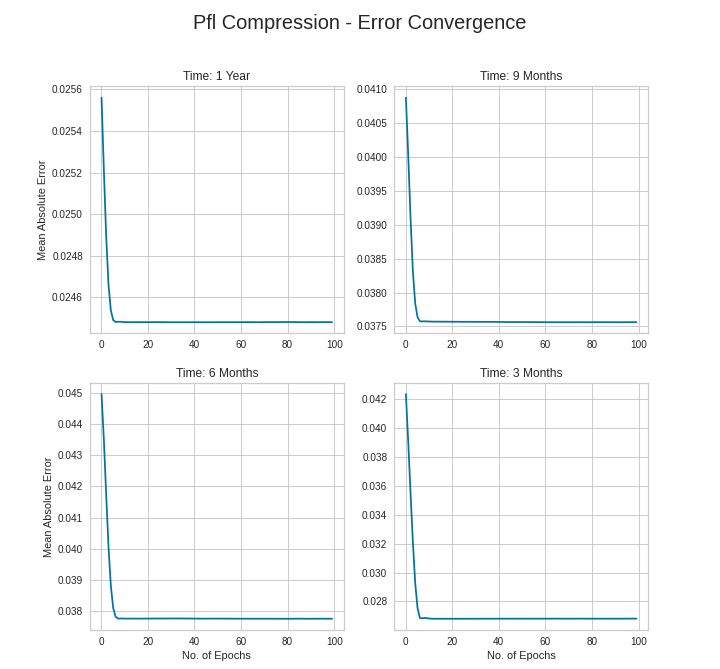}
\caption{ The plot shows the neural network model error convergence in pricing 10,000 Target Options Portfolio using 16 Compressed Options Portfolio. The x-axis corresponds to number of epochs and the y-axis corresponds to Mean Absolute Error. Each subplot corresponds to a specific time horizon (1 Year, 9 Months, 6 Months and 3 Months).}
 \label{Fig_error_convergence_100_16_10000.png}
\end{center}
\end{figure}

\noindent In this section, we illustrate the convergence of model error and evolution of neural network parameters ($K$ and $W$) across each epoch in the proposed iterative algorithm. The value of the target portfolio value $V\big(t, S_t(\omega_j)\big)$ and compressed portfolio value $P\big(t, S_t(\omega_j)\big)$ is calculated for each simulated path $\omega_j$ $\forall j =1, 2,..., N$ and at each analysis time horizon $t \in \{0.25, 0.5,$ $0.75, 1\}$ using the validation data. The model error at each time $t$ is defined as the absolute mean difference between the target and compressed portfolio value across all paths; equivalently, $ \frac{1}{N} \ \sum_{j=1}^{N} |V\big(t, S_t(\omega_j)\big) - P\big(t, S_t(\omega_j)\big)|$, where $|.|$ stands for modulus. In Figure \ref{Fig_error_convergence_100_16_10000.png}, we could observe that model error converges at less than $10$ epochs and remains consistent post that. It is noted that stopping criteria is intentionally not used for this analysis to observe the error convergence across all 100 epochs. \\

\begin{figure}[!htb]
\begin{center}
\includegraphics[width=\textwidth, height=6in]{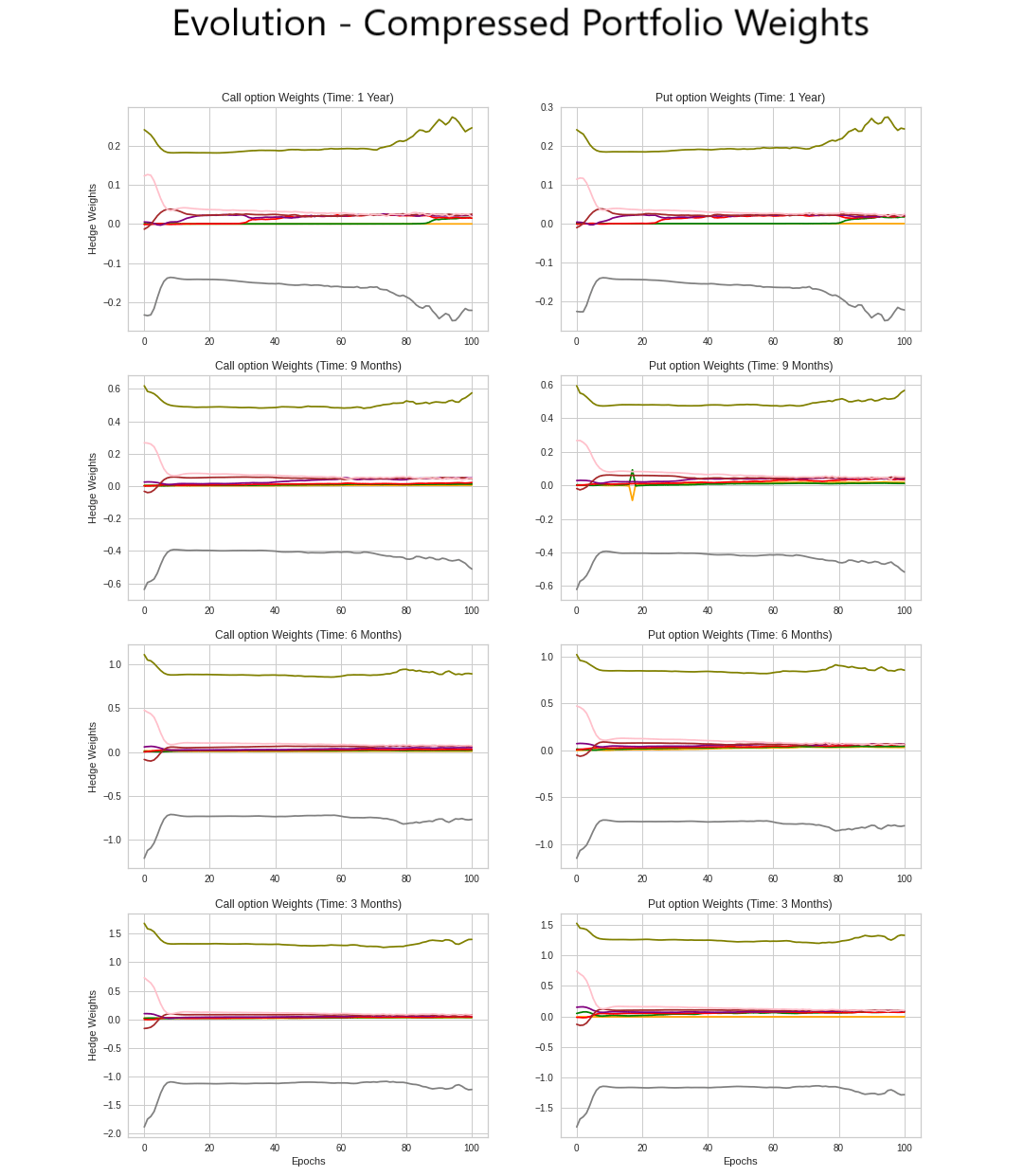}
\caption{The plot shows the evolution of the compressed options (8 calls and 8 puts) portfolio weights to price 10,000 target options portfolio. The x-axis corresponds to number of epochs and y-axis corresponds to portfolio weights value. Each row corresponds to a specific time horizon (1 Year, 9 Months, 6 Months and 3 Months). The left column corresponds to call options and right side corresponds to put options.} \label{Fig_weights_convergence_100_16_10000.png}
\end{center}
\end{figure}

\begin{figure}[!htb]
\begin{center}
\includegraphics[width=\textwidth, height=5in]{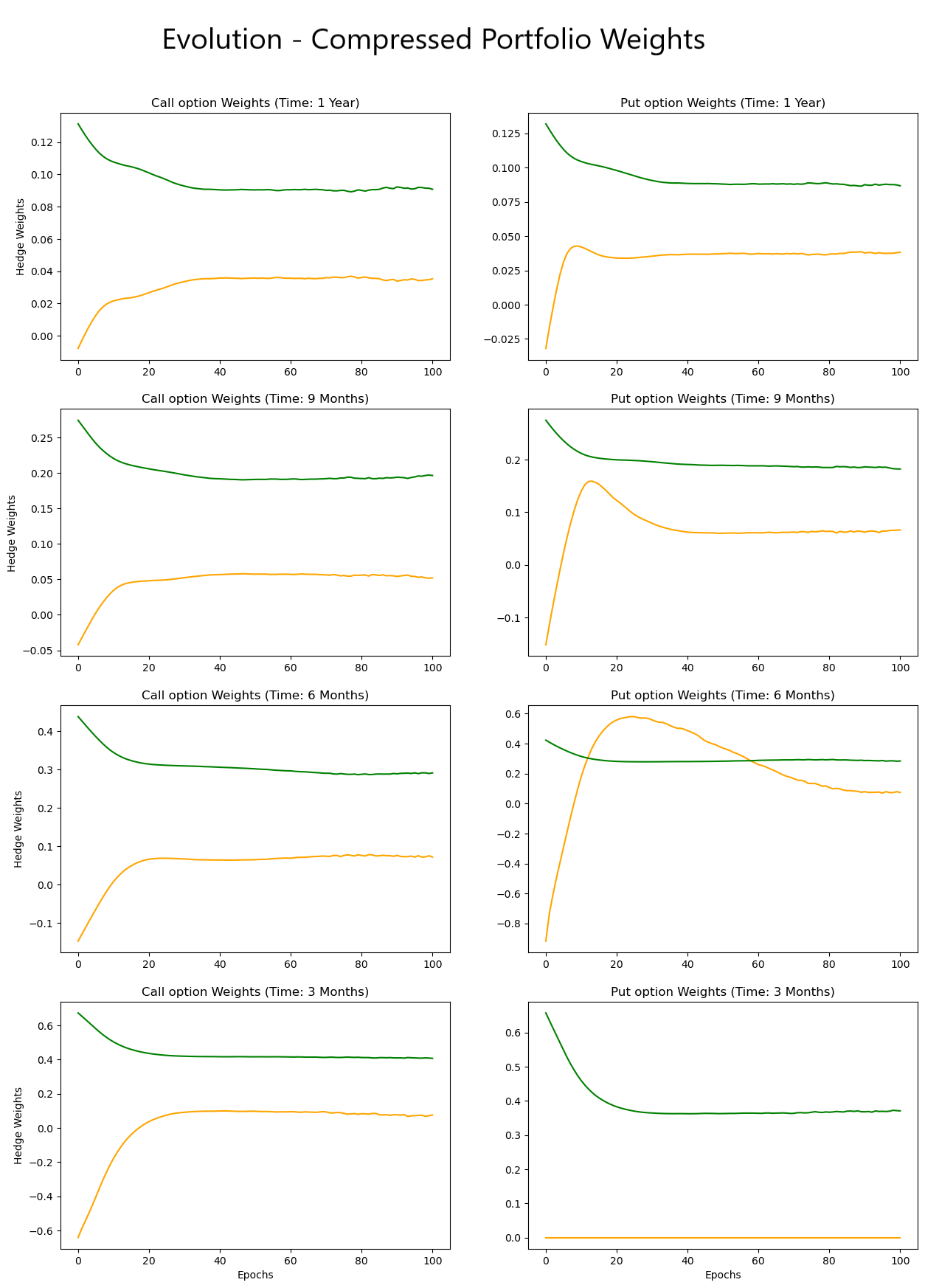}
\caption{The plot shows the evolution of the compressed options (2 calls and 2 puts) portfolio weights to price 10,000 target options portfolio. The x-axis corresponds to number of epochs and y-axis corresponds to portfolio weights value. Each row corresponds to a specific time horizon (1 Year, 9 Months, 6 Months and 3 Months). The left column corresponds to call options and right side corresponds to put options.} \label{Fig_weights_convergence_100_4_10000.png}
\end{center}
\end{figure}

\begin{figure}[!htb]
\begin{center}
\includegraphics[width=\textwidth, height=6in]{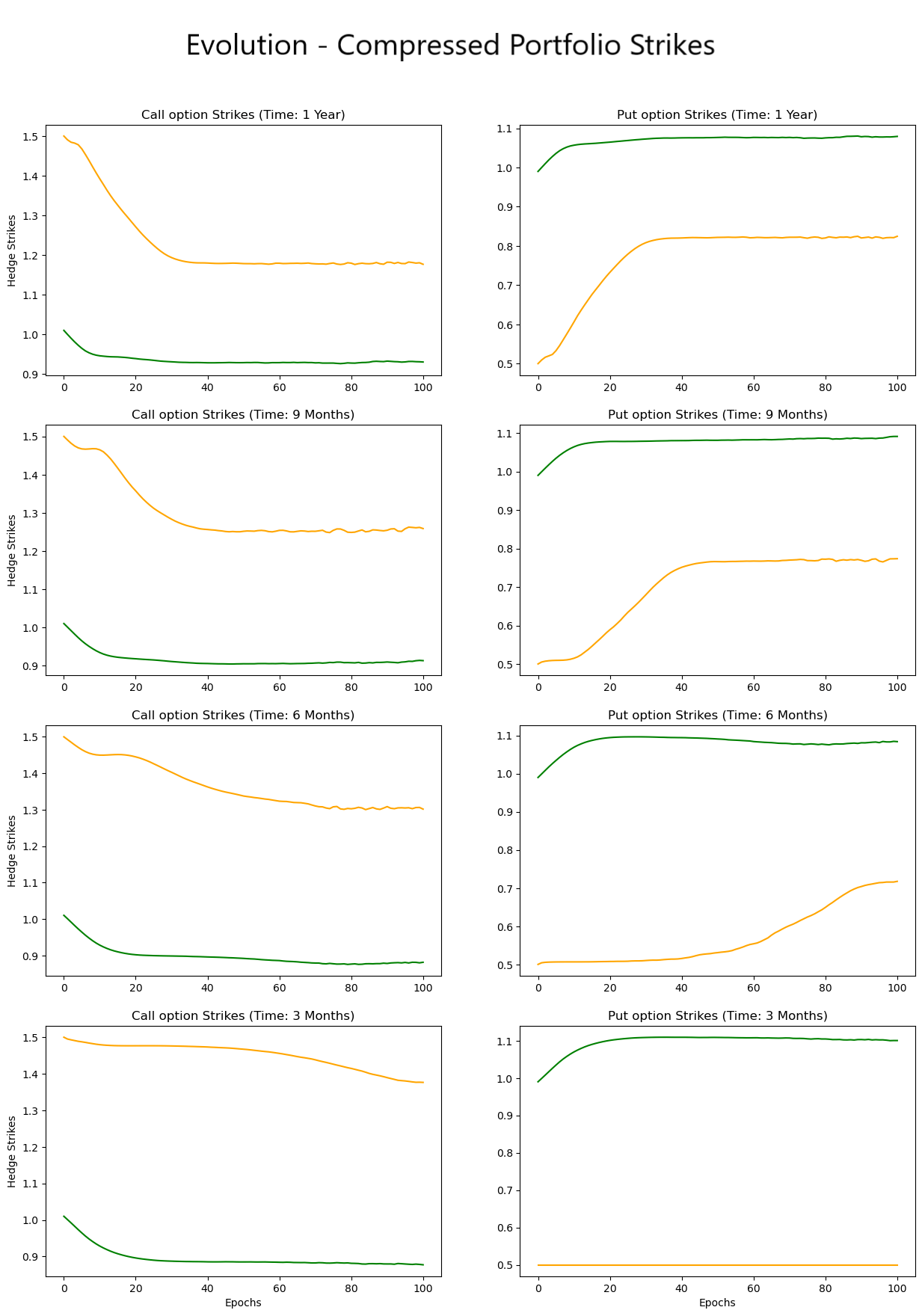}
\caption{The plot shows the evolution of the compressed options (2 calls and 2 puts) portfolio strikes to price 10,000 target options portfolio. The x-axis corresponds to number of epochs and y-axis corresponds to strike value. Each row corresponds to a specific time horizon (1 Year, 9 Months, 6 Months and 3 Months). The left column corresponds to call options and right side corresponds to put options.} 
\label{Fig_strikes_convergence_100_4_10000.png}
\end{center}
\end{figure}

\noindent In Figure \ref{Fig_weights_convergence_100_16_10000.png}, we could see the learning path of all portfolio weights for call and put options separately. There are two call options and two put options, which dominate all constituent options. Based on this observation, we also considered a case with only four constituent options in the compressed portfolio. In Figure \ref{Fig_weights_convergence_100_4_10000.png}, we observe the evolution of the four options' portfolio weights.  Figure \ref{Fig_strikes_convergence_100_4_10000.png} demonstrates the evolution of the corresponding four constituent options' strikes across epochs at different risk horizons. It is generally observed that call and put options close to ATM and OTM region contribute to the compressed portfolio. This is favourable as ATM and OTM European options are acknowledged as highly liquid instruments in the market. The accuracy and performance of compressed portfolio is studied in Section \ref{exposure_bench}.

\clearpage
\subsection{Exposures and Greeks Benchmarking}
\label{exposure_bench}

In this Section, the distribution of PV (Present Value) and Greeks (Delta, Gamma, and Vega) across all simulated paths at each time horizon, along with exposure profiles generated by risk-neutral and real-world measures, is benchmarked for the compressed and target portfolios. The PV of a portfolio in a particular simulated scenario aligns with the exposure since the constituent options are European options, devoid of path-dependency and early-exercise features.

\subsubsection{Exposures Benchmarking}

\begin{figure}[!htb]
\begin{center}
\includegraphics[width=\textwidth, height=6in]{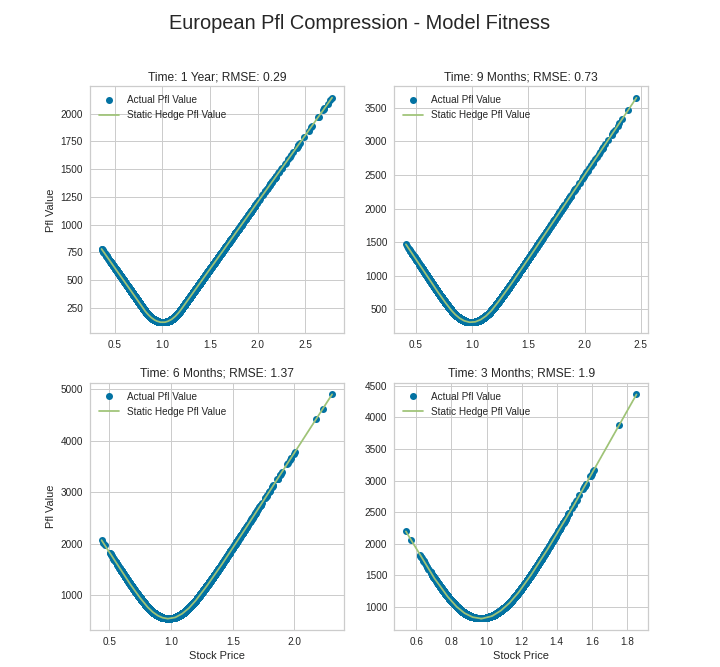}
\caption{The plot shows the distribution of PV or Exposure of 10,000 target options portfolio against 16 options compressed (or static hedge) portfolio across all simulated underlying levels under risk-neutral measure. Each subplot corresponds to one specific time horizon (1 Year, 9 Months, 6 Months and 3 Months). The x-axis corresponds to underlying stock levels and y-axis corresponds to to PV.} \label{Fig_pv_dist_100_16_10000.png}
\end{center}
\end{figure}

\begin{figure}[!htb]
\begin{center}
\includegraphics[width=\textwidth, height=3in]{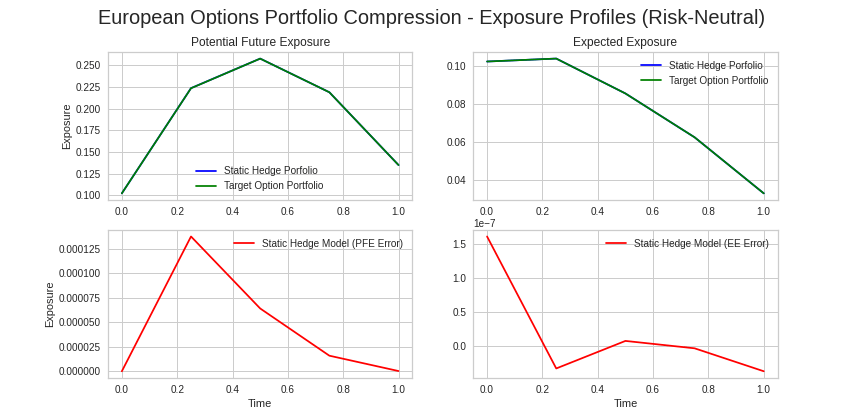}
\caption{The plot compares the EE (right) and PFE (left) of the target and compressed (or static hedge) portfolio calculated under risk-neutral measure. The target portfolio constituted 10,000 options and compressed portfolio is constructed with 16 options. The x-axis corresponds to time horizon and y-axis corresponds to exposures.} \label{Fig_exposures_riskneutral_100_16_10000.png}
\end{center}
\end{figure}

\noindent In this analysis, we create the exposure distribution at each analysis risk horizon $t \in \{0.25, 0.5, 0.75, 1\}$ for the target portfolio $V\big(t, S_t\big)$  and compressed portfolio $P\big(t, S_t\big)$ by Monte-Carlo simulation (the validation data) of the underlying under risk-neutral measure and revaluation of the portfolio at each simulated levels. In Figure \ref{Fig_pv_dist_100_16_10000.png}, we compare the exposure distribution of the target portfolio and compressed portfolio (with 16 constituent options). The maximum Root Mean Square Error (RMSE) as the ratio of the number of target options across all risk horizons is in the order of $10^{-3}$. When a compressed portfolio with only four options, we observed the ratio is maximum at $10^{-2}$. As expected, performance increases with the number of constituent options in the compressed portfolio. The selection of optimal options is the principal factor driving the balance between accuracy and compressed portfolio size (number of constituent option strikes). The Expected Exposure (EE) profile and $99\%$ Potential Future Exposure (PFE) profile are computed. The EE for the target and compressed portfolio are $\frac{1}{N} \cdot \sum_{j=1}^{N} P\big(t, S_t(\omega_j)\big)$ and $\frac{1}{N} \cdot \sum_{j=1}^{N} V\big(t, S_t(\omega_j)\big)$ respectively. The PFE for the target and compressed portfolio are $99^{th}$ percentile of $V\big(t, S_t\big)$ and $P\big(t, S_t\big)$ respectively for each risk horizon $t$. Exposures Error is the difference between target and compressed portfolio exposures for each risk horizon. In Figure \ref{Fig_exposures_riskneutral_100_16_10000.png}, we observe that the exposure profiles (scaled down by number of target options $M$) of target and compressed portfolio align each other with PFE error and EE error in the order of $10^{-4}$ and $10^{-7}$ respectively. Based on the experiments with four options case, PFE and EE errors were observed in the order of $10^{-3}$ and $10^{-4}$. Overall, a good alignment of exposure profiles between the target and compressed portfolio is observed. \\

\noindent The four stressed time zero market data scenarios have been generated and are considered real-world market data scenarios to benchmark the exposure profiles of the compressed portfolio against the target portfolio. The four real-world time-zero market scenarios are:

\begin{itemize}
\item Scenario 1: $S_0 =1 , \sigma_{real}=0.1, \mu=0.07$
\item Scenario 2: $S_0 =1 , \sigma_{real}=0.3, \mu=0.1$
\item Scenario 3: $S_0 =1 , \sigma_{real}=0.5, \mu=0.15$
\item Scenario 4: $S_0 =1 , \sigma_{real}=0.5, \mu=0.01$
\end{itemize}

\noindent The GBM process generates the future underlying distribution with assumed underlying returns $\mu$ and volatility $\sigma_{real}$ corresponding to the four market scenarios. Then, the two portfolios are re-valued at each simulated path across all risk horizons for each market scenario. In other words, the underlying is simulated under real-world measure and the two portfolios are valued under risk-neutral measure. As a result, the EE and PFE profiles of the target and compressed portfolio are obtained.  The profiles were observed to have a good alignment for all four real-world scenarios considered, as observed in the Appendix Section \ref{16opt_real_exp}.

\subsubsection{Greeks Benchmarking}
\begin{figure}[!htbp]
\begin{center}
\includegraphics[width=0.9\textwidth, height=4.7in]{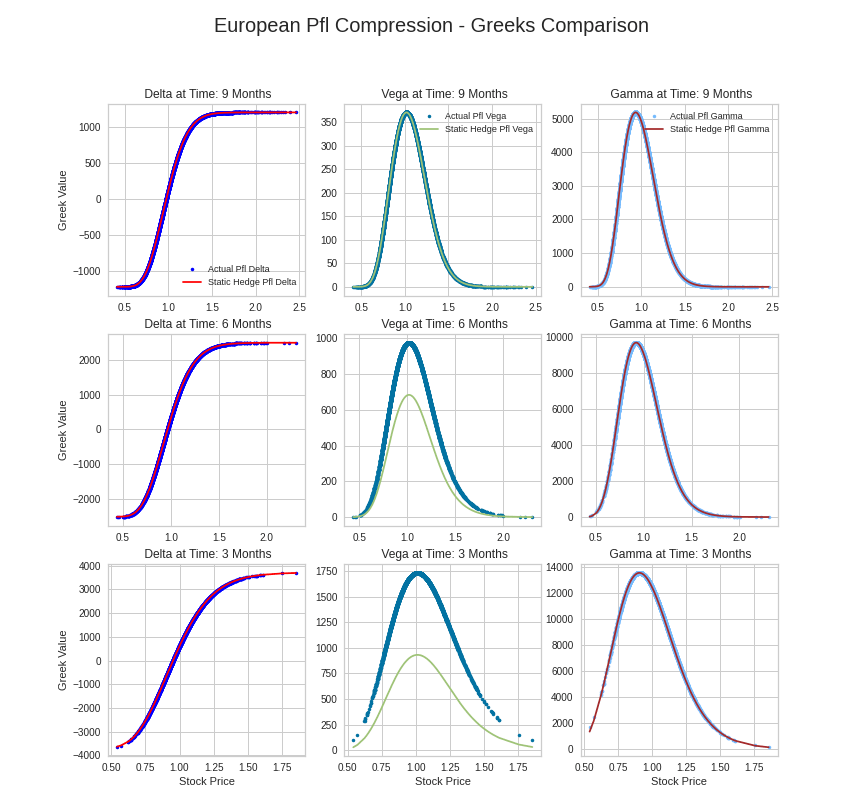}
\caption{The plot compares the Greeks of the target portfolio of 10,000 options and compressed portfolio of 16 options. The delta, gamma and vega are compared in the first, second and third columns respectively. Each row corresponds to specific time horizon (3 Months, 6 Months and 9 Months). The x-axis corresponds to simulated underlying stock and y-axis corresponds to the Greek value.} \label{Fig_greeks_dist_100_16_10000.png}
\end{center}
\end{figure}

\noindent In this section, we compare the option sensitivities/Greeks, namely Delta, Gamma, and Vega of the target portfolio and compressed portfolio as illustrated in Figure \ref{Fig_greeks_dist_100_16_10000.png}. The portfolio level Greeks are obtained as the dot product of respective sensitivity vector of component options and portfolio weights. At all risk horizons and simulation paths, Delta and Gamma of target and compressed portfolios are consistent, whereas Vega aligns only at nine months horizon. The compressed portfolio's Vega is predominantly lower at three and six months risk horizons, which is more evident in the ATM region. This is because, at all risk horizons, the maturity of a compressed portfolio is three months. In contrast, at three-months risk horizon, the target portfolio comprises three, six, and nine months maturity options. Similarly, at six-months risk horizon, the target portfolio comprises options expiring in three and six months. The difference in maturities of constituent options between the target and compressed portfolios drives this vega difference. At nine months risk horizon, both portfolios have the same maturity leading to consistent Vega. If there is a requirement to match the Vega, it is necessary to construct a compressed portfolio with the same maturity as the target portfolio. However, Vega does not impact the Profit and Loss of the portfolio under a constant volatility framework, which is the scope of this paper.

\subsection{Standardised Capital}
\label{Standardised Capital}

In this section, the Exposure At Default (EAD) based on a standardised approach for the counterparty credit risk as per the BASEL norms in \cite{biscre52} is calculated for both target (with 10,000 options) and compressed portfolio (with 16 options) for comparison. \\

\noindent We assume that the analysis portfolios (target and compressed) are un-margined and have zero initial collateral. In addition, we also assume that the target and compressed portfolio comprises only vanilla equity single-name options. As a result, from the Basel framework, supervisory volatility $\sigma_{sup}$ and supervisory factor $SF$ are : 
\begin{align*}
 \sigma_{sup} &= 120\%, \\
  SF &= 32\%.
 \end{align*}
The supervisory volatility and supervisory factor are used in the EAD calculation detailed below. The standardised capital methodology applicable for the test portfolio as per BASEL norms is summarised below:

\begin{align} 
EAD &= 1.4 \cdot (RC + PFE), \\
RC &= max\{Q-C, 0\}, \\
PFE &= multiplier \cdot AddOn^{aggregate}, \label{std_cap}
\end{align}

\noindent where, $RC$ is Replacement Cost, \\ 
$PFE$ is Potential Future Exposure, \\
$Q=V(t_0, S_0)$ for the target portfolio and $Q=P(t_0, S_0)$ for the compressed portfolio, both are expected to be close by the design of the algorithm, \\
$C$ is the post-haircut value of the net initial collateral, which is assumed to be zero for the analysis. \\

\noindent In Equation \ref{std_cap}, 
\begin{align}
multiplier &= min\{1, \ 0.05 + 0.95 \cdot exp\Big(  \frac{Q-C}{2 \cdot 0.95 \cdot AddOn^{aggregate}} \Big)\}, \\
AddOn^{aggregate} &= \sum_{asset \ class} AddOn^{(asset \ class)}, \\
AddOn^{(asset \ class)} &= \Big[ \big( \sum_{entity} \rho_{entity} \cdot AddOn^{(entity)} \big)^{2}  \nonumber \\ 
& \ \ \ \ \ + \sum_{entity} (1 - \rho^{2}_{entity} ) \cdot ( AddOn^{(entity)} )^{2} \Big]^{0.5}.
\end{align}

\noindent In our analysis, we consider only one asset class; thereby, $AddOn^{aggregate}$ and  $AddOn^{(asset \ class)}$ are the same.  $AddOn^{(entity)}$ is the sum of Effective Notional ($EN(i)$) of each each constituent option $i$ for an entity, 
\begin{align}
AddOn^{(entity)} = \sum_{i} EN(i) \cdot SF.
\end{align}

\noindent For each constituent option $i$ of the portfolio of interest with corresponding portfolio weight $w_i$, 

\begin{equation} \label{EN}
EN(i) = AN(i) \cdot SD(i) \cdot MF(i),
\end{equation}
where, $AN(i)$ is Adjusted Notional, \\ 
$SD(i)$ is Supervisory Delta,\\ 
$MF(i)$ is Maturity Factor. \\

\noindent Further in Equation \ref{EN}, 

\begin{align}
AN(i) &= |w_i| \cdot S_0, \\
SD(i) &= i_{ls} \cdot \Phi\Big(i_{cp} \cdot \frac{ln\big(\frac{S_0}{k(i)}\big) + 0.5 \cdot \sigma_{sup}^{2} \cdot maturity(i)}{\sigma_{sup} \cdot \sqrt{maturity(i)}}\Big), \\
MF(i) &= \sqrt{\frac{maturity(i), 1 \ year}{1 \ year}}
\end{align}

\noindent where,\\  
$i_{ls} = +1$ for long call and short put position, whereas, $i_{ls} = -1$ for short call and long put position, \\ 
$i_{cp}=+1$ for call option and $i_{cp}=-1$ for put option, \\ 
$\Phi(\cdot)$ correspond to Standard Normal CDF function, \\
$maturity(i)$ refers to the maturity of the constituent option $i$ and can take values $\{0.25, 0.5, 0.75, 1.0\}$ for target options portfolio, whereas, maturity is always $0.25$ for the compressed portfolio, \\
$k(i)$ corresponds to the strike of the constituent option of the portfolio of interest, \\
$S_0$ is the time zero spot price. \\

\begin{table}[htbp]
  \centering
\resizebox{\textwidth}{!} {
    \begin{tabular}{|c|c|c|c|c|c|} 
    \hline 
    \noalign{\vspace{2pt}}
    \multicolumn{1}{|c|}{\multirow{2}[4]{*}{\textbf{S.No.}}} & \multicolumn{2}{p{12.2755em}|}{\centering \textbf{Target Portfolio}} & \multicolumn{2}{p{11.545em}|}{\centering \textbf{Compressed Portfolio}} & \multicolumn{1}{c|}{\multirow{2}[4]{*}{\textbf{Reduction in EAD (\%)}}} \\
\cmidrule{2-5}          & \textbf{Portfolio Composition} & \multicolumn{1}{p{5.5955em}|}{\textbf{EAD (RC)}} & \textbf{Portfolio Composition} & \multicolumn{1}{p{5.955em}|}{\textbf{EAD (RC)}} &  \\
    \midrule
    1     & 10,000 Long Options:  &         2,071 (1,022)  & 16 Options:  &          1,721 (1,022)  & 16.9\% \\
         & 5000 Calls + 5000 Puts &           & 8 Calls + 8 Puts  &            &  \\
    \midrule
    2     & 10,000 Long Call Options  &         4,031 (1,181)  & 16 Options:  &      3,060 (1,181)     & 24.1\% \\
        &   &            & 8 Calls + 8 Puts  &         &   \\
    \midrule
    3     & 10,000 Long Put Options  &            2,291 (872)  & 16 Options:  &    2,015 (872)  & 12\% \\
         & &             & 8 Calls + 8 Puts  &               &  \\
\midrule
    4     & 10,000 Long \& Short Options:  &        1,602 (523)     & 16 Options:  &     1,286 (523)    &  19.7\% \\
         & 5000 Calls + 5000 Puts  &             & 8 Calls + 8 Puts  &               &   \\[5pt]
    \hline
    \end{tabular}%
}
 
  \caption{The table compares EAD and RC of four different target portfolios with their corresponding compressed portfolios. The constituent options of the target portfolio are selected with maturities 3 months, 6 months, 9 months and one year uniformly. The compressed portfolio expires in 3 months. The reduction in EAD and therefore capital is observed for compressed portfolio.}
  \label{tab:standardised_capital}%
\end{table}%

\begin{table}[htbp]
  \centering
\resizebox{\textwidth}{!} {
    \begin{tabular}{|c|c|c|c|c|c|} 
    \hline 
    \noalign{\vspace{2pt}}
    \multicolumn{1}{|c|}{\multirow{2}[4]{*}{\textbf{S.No.}}} & \multicolumn{2}{p{12.2755em}|}{\centering \textbf{Target Portfolio}} & \multicolumn{2}{p{11.545em}|}{\centering \textbf{Compressed Portfolio}} & \multicolumn{1}{c|}{\multirow{2}[4]{*}{\textbf{Reduction in EAD (\%)}}} \\
\cmidrule{2-5}          & \textbf{Portfolio Composition} & \multicolumn{1}{p{5.5955em}|}{\textbf{EAD (RC)}} & \textbf{Portfolio Composition} & \multicolumn{1}{p{5.955em}|}{\textbf{EAD (RC)}} &  \\
    \midrule
    1     & 10,000 Long Options:  &         1,357 (757)  & 16 Options:  &  1,358 (757)  & 0.07\% \\
         & 5000 Calls + 5000 Puts &          & 8 Calls + 8 Puts  &           &  \\
    \midrule
    2     & 10,000 Long Call Options  &     3,919 (819)      & 16 Options:  &     3,921 (819)     & 0.05\% \\
        &   &            & 8 Calls + 8 Puts  &         &   \\
    \midrule
    3     & 10,000 Long Put Options  &            2,596 (695)   & 16 Options:  &    2,605 (695)   & 0.3\% \\
         & &             & 8 Calls + 8 Puts  &               &  \\
\midrule
    4     & 10,000 Long \& Short Options:  &        1,243 (376)     & 16 Options:  &     1,249 (376)    &  0.4\% \\
         & 5000 Calls + 5000 Puts  &             & 8 Calls + 8 Puts  &               &   \\[5pt]
    \hline
    \end{tabular}%
}
  \captionsetup{skip=10pt}
  \caption{The table compares EAD and RC of four different target portfolios with their corresponding compressed portfolios. The constituent options of the target portfolio and the compressed portfolio are of same maturity. The EAD is close for the target and the compressed portfolios.}
  \label{tab:standardised_capital-samemat}%
\end{table}%

\noindent We consider four target portfolios to compare the EAD by standardised approach. In all four cases, we have 10,000 constituent options, where, in the first and last case, it's a combination of 5000 calls and 5000 puts, while in the second case, it's all call options, whereas in the third case, only put options. In the last case, we have both long and short positions, but other cases comprise only long positions. It is to be noted that each option in the target portfolio of the last case is assigned randomly as buy and sell with 75\% and 25\% probability, respectively. Table \ref{tab:standardised_capital} shows that the compressed portfolio's capital is significantly reduced by $16.9\%$, $24.1\%$, $12\%$, $19.7\%$ for the test portfolios in the same order. \\

\noindent This reduction in the capital is primarily because the compressed portfolio is constructed as a shorter maturity portfolio replicating the target portfolio value at future time points. In other words, we can consider this as breaking the longer-term target portfolio into self-replicating short-term portfolios, which, therefore, can be re-invested after each shorter maturity interval to replicate the target portfolio. Hence, the Maturity Factor would be lesser for the compressed portfolio, which is critical in reducing the capital. In Table \ref{tab:standardised_capital-samemat}, we can observe that the EAD of target and static portfolios closely align as the constituent options of both portfolios are of the same maturity (3 months). It was also observed that both portfolios' supervisory delta closely align. This adds more evidence to show that the shorter maturity of a compressed portfolio is the driving factor for reduced capital. Further, if the target portfolio comprises long and short positions, the algorithm performs an auto netting as the netted target portfolio value $V(t, S_t)$ would be used in the model $NN(t)$ at all $t$ to construct the compressed portfolio and supplemented with an additional advantage of reduced maturity factor for capital calculation.

\section{Conclusion}
In this chapter, we extended the application of the neural network architecture (introduced in \cite{dhandapani2024bermudan}) with defined model parameters initialisation and optimisation algorithms to build a compressed static hedge portfolio for a large portfolio of target European options. We also illustrated through numerical examples the capability of the algorithm to build a significantly small compressed portfolio for a large target portfolio and still achieve a close alignment of Exposure distribution and profiles of the portfolios under both risk-neutral and presumed real-world scenarios. We also benchmarked the portfolio Greeks at all future simulated risk-neutral scenarios. We observed a close alignment of Delta and Gamma. In contrast, the compressed portfolio's Vega was lower than the target portfolio's Vega if the maturity of the compressed portfolio is lesser than the target portfolio. The difference is more evident as we move towards the ATM region. If they are of the same maturity, the Vega profiles agree. However, it is noted that there is no vega exposure in the constant volatility framework, which is under the scope of this paper. Finally, we demonstrated with numerical examples that the standardised capital charge for the compressed portfolio's counterparty credit risk exposure is significantly reduced compared to the target portfolio primarily because of the low PFE driven by the portfolio characteristics - maturity factor and portfolio composition. \\ 

\noindent The proposed portfolio compression or static hedging framework is compatible with only the constant volatility market environment, and adapting to the stochastic volatility environment is a potential future scope.

\section*{Disclosure statement}

The authors report there are no competing interests to declare.

\section*{Funding}

The authors declare that no funds, grants, or other support were received during the preparation of this manuscript.

\bibliographystyle{tfs.bst}
\bibliography{references}

\clearpage

\appendix

\section{Compressed Portfolio Exposure Profiles (16 Constituent options) under real-world scenarios} \label{16opt_real_exp}

\begin{figure}[!htb]%
 \centering
 \subfloat[Scenario 1]{\includegraphics[width=0.5\textwidth, height=1.5in]{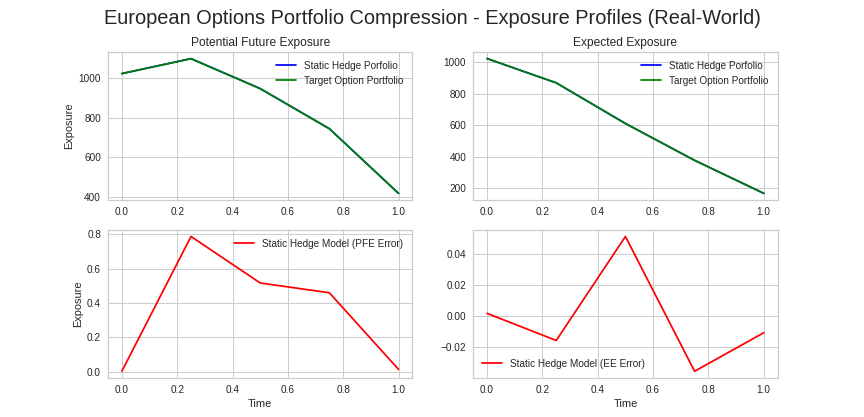}\label{Fig_exposures_realworld_100_16_10000_scen1.png}}%
 \subfloat[Scenario 2]{\includegraphics[width=0.5\textwidth, height=1.5in]{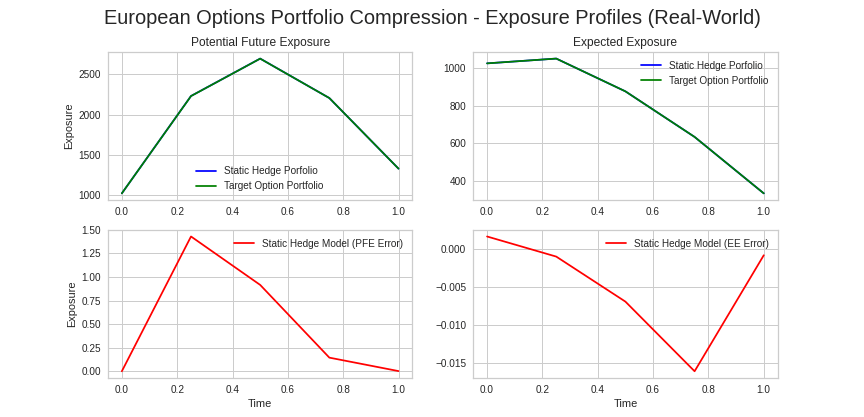}\label{Fig_exposures_realworld_100_16_10000_scen2.png}}\\
 \subfloat[Scenario 3]{\includegraphics[width=0.5\textwidth, height=1.5in]{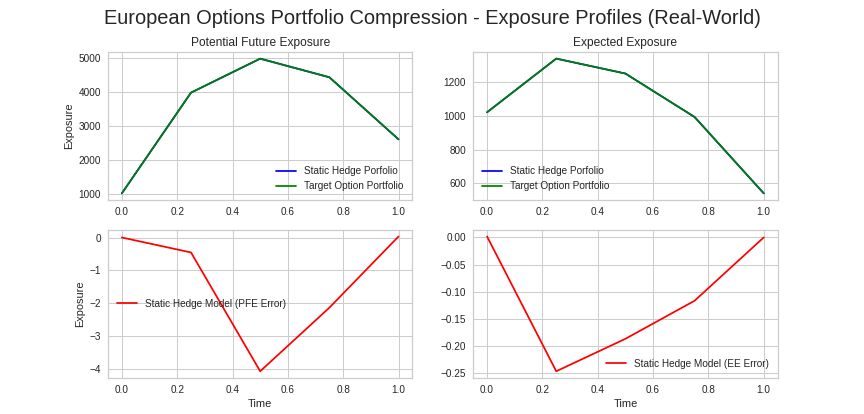}\label{Fig_exposures_realworld_100_16_10000_scen3.png}}%
 \subfloat[Scenario 4]{\includegraphics[width=0.5\textwidth, height=1.5in]{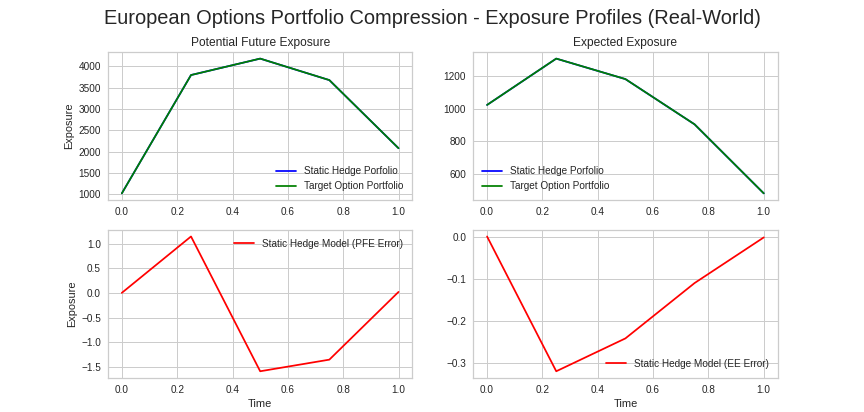}\label{Fig_exposures_realworld_100_16_10000_scen4.png}}%
 \caption{The plot compares the EE and PFE of the target and compressed (or static hedge) portfolio calculated under real-world scenarios. The target portfolio constituted 10,000 options and compressed portfolio is constructed with 16 options. The x-axis corresponds to time horizon and y-axis corresponds to exposures. The four subplots correspond to the following four time-zero market data scenarios: Scenario 1: $S_0 =1 , \sigma_{real}=0.1, \mu=0.07$; Scenario 2: $S_0 =1 , \sigma_{real}=0.3, \mu=0.1$; Scenario 3: $S_0 =1 , \sigma_{real}=0.5, \mu=0.15$; Scenario 4: $S_0 =1 , \sigma_{real}=0.5, \mu=0.01$;}%
\label{EE_Pfl_real_world}%
\end{figure}

\end{document}